\documentclass{gretsi}

\usepackage[english,french]{babel}   % "babel.sty" + "french.sty"
\usepackage{times}			% ajout times le 30 mai 2003
 \usepackage{graphicx}
\title{Regression Constraint for an Explainable Cervical Cancer Classifier\\}

\author{\coord{Antoine}{Pirovano}{1,2},
        \coord{Leandro G.}{Almeida}{1},
    \coord{Said}{Ladjal}{2}}

\address{\affil{1}{Keen Eye,
         74 Rue du Faubourg Saint-Antoine, 75012 Paris, France}
         \affil{2}{Laboratoire Traitement et Communication de l'Information, Telecom ParisTech, Universite Paris-Saclay, Paris, France\\}}

\email{antoine.pirovano@keeneye.tech, leandro.almeida@keeneye.tech, said.ladjal@telecom-paristech.fr}

\frenchabstract{Cet article s'int\'{e}resse \`{a} la classification automatique de cellules de l'\'{e}pith\'{e}lium pavimenteux pour le depistage cancer du col de l'ut\'{e}rus en s'appuyant sur les outils de l'apprentissage profond. Nous \'{e}tudions differentes architectures sur un jeu de donn\'{e}es public nomm\'{e} Herlev qui consiste \`{a} classifier des images de cellules, issues d'un frottis du col de l'ut\'{e}rus, au regard de l'anormalit\'{e} qu'elles repr\'{e}sentent. De plus, nous utilisons et adaptons une m\'{e}thode d'attribution afin de mettre en lumi\`{e}re les carat\'{e}ristiques cytomorphologiques discriminantes qui sont utilis\'{e}es pour la classification. A travers cet article, nous d\'{e}taillerons les methodes et architectures qui nous permettent d'atteindre des performances optimis\'{e}es: 74.5\% de pr\'{e}cision pour la classification de la s\'{e}v\'{e}rit\'{e} et 94\% pour la classification de la normalit\'{e}.}

\englishabstract{This article adresses the problem of automatic squamous cells classification for cervical cancer screening using Deep Learning methods. We study different architectures on a public dataset called Herlev dataset, which consists in classifying cells, obtained by cervical pap smear, regarding the severity of the abnormalities they represent. Furthermore, we use an attribution method to understand which cytomorphological features are actually learned as discriminative to classify severity of the abnormalities. Through this paper, we show how we trained a performant classifier: 74.5\% accuracy on severity classification and 94\% accuracy on normal/abnormal classification.}

\begin{document}
\maketitle

\section{Introduction}
\label{intro}
The World Health Organization (WHO) states \cite{WHO06} that around 90\% of
cervical cancer could be avoided if they were detected and treated earlier.
At $500\times 10^3$ new cases per year, screening for cervical
cancer needs to be efficient and precise.

With the recent emergence of machine learning using deep Convolutional
Neural Networks (CNN) and its success on a large panel of tasks, a
lot of work has been done to assist doctors and medical practices
\cite{Bar15,Ronneberger15} using such methods. In the case of cervical
cancer, the Herlev public dataset enables to compare different methods
on this specific task by providing images of single cells and organizing
them into classes regarding the malignancy they represent.

In this paper, we will firstly exploit the ordinal nature of the WHO classification
present in the Herlev dataset, by designing a loss function that leads to a
training paradigm that closely resembles the medical task at hand. Finally we will
apply attribution methods to determine what cytomorphological features are associated
with the classification model. This will not only give us confidence in the
training process and prove that the model learned relevant features
but also show the potential for weak localization tasks.

\section{Related Work}
\label{related}
Since 2012 and the success of AlexNet on Imagenet Challenge
\cite{Krizhevsky12}, deep CNN have provided high accuracy results in
 large range of different tasks. Over the years,
several architectures have been given a lot of attention. For example,
Resnet-101 \cite{He15} proposes to use skipped connections over blocks
to avoid unlearning on more abstract features spaces.

Previous works have applied CNN models to the Herlev data set using binary normal
and abnormal categories. In \cite{Hyeon17} they reach a 0.78 F1 scoring using a
support vector machine. In \cite{Bora16} they use a unsupervisely trained Feature
Selection model after a CNN feature extractor to reach a F1 score of 0.90 and an
accuracy of 94\%. In \cite{Lu17, Forslid17}, they used, respectively, an
Alexnet-like and a Resnet architecture and trained them on Herlev dataset using
normal vs abnormal to provide a model that reaches binary classification
accuracy of 98.3\%.

\section{Herlev Severity Classification using Regression Constraint}
\label{improvingherlevclassif}

\subsection{Herlev Dataset}

The Herlev Dataset is a cytology image set composed of 917 images gathered in 7
classes:  normal columnar, normal intermediate, normal superficial, light
dysplastic, moderate dysplastic, severe dysplastic, and carcinoma in situ. The
three first classes belong to the category of normal cells and the last four are
abnormal ones (in order of severity, carcinoma in situ hinting at the presence
of an actual cancer). Images are between 50 and 400 pixels wide. Previous work
processed the set in a binary classification problem of normal vs abnormal
classes. Here, we merged normal images into a single class in order to study the
medical severity only, thus building a 5 classes dataset, we call Herlev
severity consisting in : normal, light dysplastic, moderate dysplastic,
severe dysplastic and carcinoma in situ.

\subsection{Herlev Severity}
\label{severityclassifiers}

This section descibes three pipelines that can be used to train a severity model
 and the motivation that led to them.

\subsubsection{Classification Pipeline}

We started by retraining a Resnet-101 model pretrained on ImageNet \cite{Deng09}
on Herlev severity dataset. The computed performances were a mean AUC of 0.9,
with the highest AUC being 0.95 on the carcinoma in situ class and lowest being
0.87 on severe dysplastic with an overall accuracy of 70.1\%, a binary
(normal/abnormal) accuracy of 90.8\% and a binary F1 score of 0.94.

\begin{figure}[!t]
\centering
\includegraphics[width=0.75\linewidth]{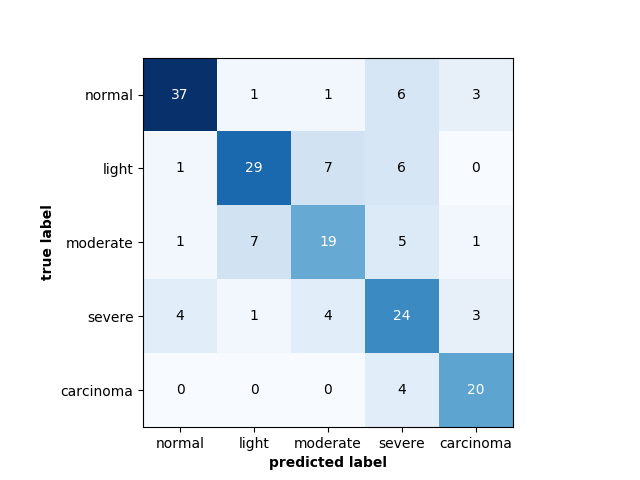}
\caption{Resnet-101 confusion matrix on Herlev severity test set}
\label{herlevsevconfmat}
\end{figure}

From the confusion matrix shown in Figure \ref{herlevsevconfmat}, we see that
the model tends to misclassify images from the normal class and most severe classes
(severe dysplastic and carcinoma in situ). This was already reported in \cite{Lu17} and identified to be
due to the visual similarities between normal columnar and carcinoma in situ cells.
Obviously, missing a potential highly abnormal diagnonis is something to avoid.
Similarly, due to the fact that 93\% of pap smears are normal during
routine diagnonis, misclassifying normal cells would require a additional action
by the attending cytotechnicians.

\subsubsection{Regression Pipeline}

Since the WHO classification used in the Herlev set have an order of severity,
this task can be interpreted as a regression problem.
Regression loss will oblige the network to focus on how to differentiate
normal samples from malignant ones. We relabel Herlev samples using a score
from 1 (for normal ones) to 5 (for carcinoma ones) and use a mean square error
as loss to optimize. Thus, we retrain the exact same Resnet-101 architecture replacing,
to have a single score output, the softmax layer by a fully connected layer.

\begin{figure}[!t]
\centering
\includegraphics[width=0.85\columnwidth]{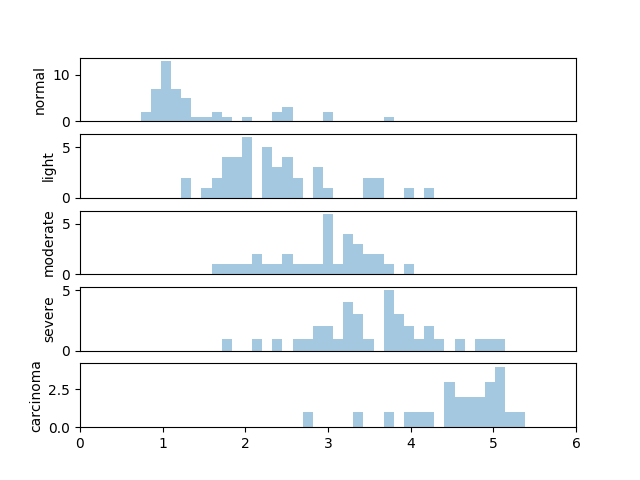}
\caption{Scores Distribution given by Resnet-101 Regressor regarding Herlev Severity Classes}
\label{herlevregressordistrib}
\end{figure}

\begin{figure}[!t]
\centering
\includegraphics[width=0.85\columnwidth]{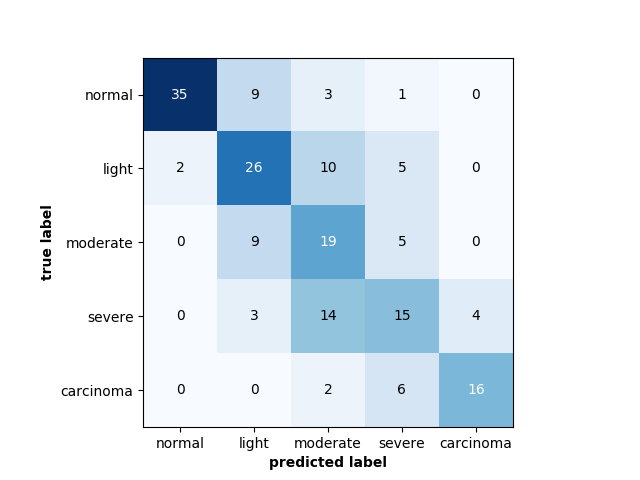}
\caption{Confusion Matrix given by Resnet-101 Regressor regarding Herlev Severity Classes}
\label{herlevregressorconfmat}
\end{figure}

Figure \ref{herlevregressordistrib} shows the distribution of scores predicted on
the test set and highlights that the model succeeded in
assigning scores regarding maligancy. Most importantly, it does not mis-classify
any normal samples or carcinoma in situ samples with each other. A further
point to note from the confusion matrix (Figure \ref{herlevregressorconfmat}) deriving from this distribution, this
model does more mis-classifications than the categorical model, with an accuracy of
60.6\%, however these misclassifcation are less severe in the scope due to their
relative prognosis distance. This is can be more easily displayed by the
overall MSE of 0.58 over the test set. The binary accuracy was of 91.8\% and the
F1 score was 0.95.

\subsubsection{{Classification + Regression} Pipeline}
\label{herlevclassifierregressor}

While the regression loss was more adapted than a classification (cross entropy)
loss to the severity task, it nonetheless did not improve the
performances per class. In this section we combine the strength of both approaches
into a single architecture.

\begin{figure}[!t]
\centering
\includegraphics[width=0.7\columnwidth]{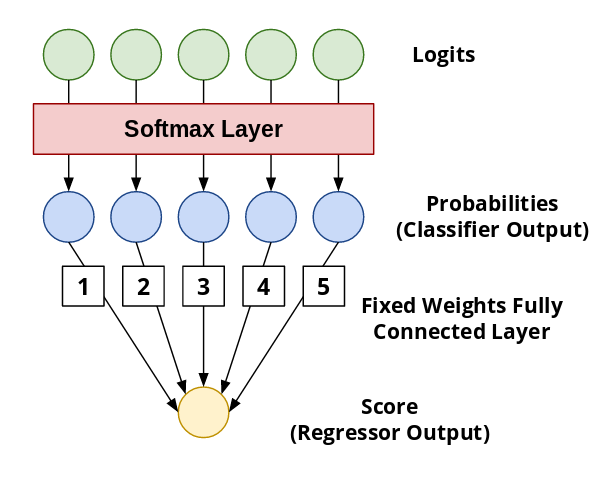}
\caption{Regression Constraint applied to Classifier}
\label{regclassarchi}
\end{figure}

Figure \ref{regclassarchi} shows the additional layer to the
classification architecture. We simply sum the cross entropy loss and the MSE
loss. This would be equivalent to weighting loss regarding the distance between
the ground truth class index and the predicted class index. We turn
probabilities given by the softmax layer into a score using a fixed weights
fully connected layer corresponding to the class score (or class index).
% A shift of 1
% on scores is made so the first class normal is represented by '1' scoring and
% not '0' to avoid gradient vanishing.

Noting $p = (p_1,\dots,p_5)$ these class probability neurons, our loss finally
reads $\mathcal{L}(x) = \mathcal{CE}(p; y_x) + (y_x-\sum_{i=0}^{4} (i+1).p_i)^2$
where $x$ is an image, $y_x$ the label (one hot for cross-entropy and score
for the regression constraint) and $\mathcal{CE}$ is the
cross-entropy loss.

\begin{figure}[!t]
\centering
\includegraphics[width=0.85\columnwidth]{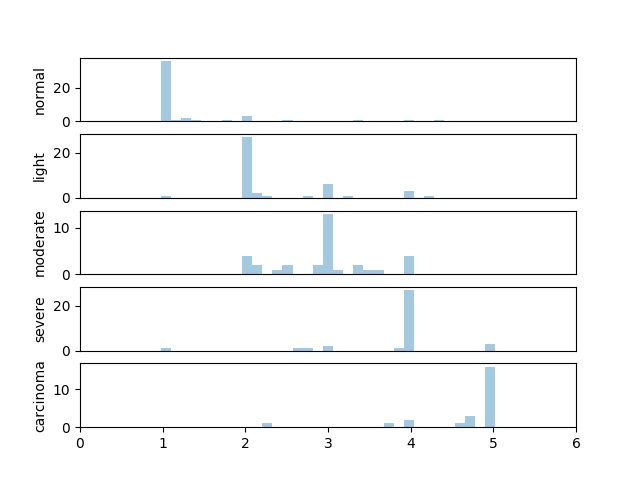}
\caption{Scores Distribution given by Resnet-101 {Classifier + Regressor} on Herlev Severity Test Set}
\label{herlevclassregdistrib}
\end{figure}

\begin{figure}[!t]
\centering
\includegraphics[width=0.85\columnwidth]{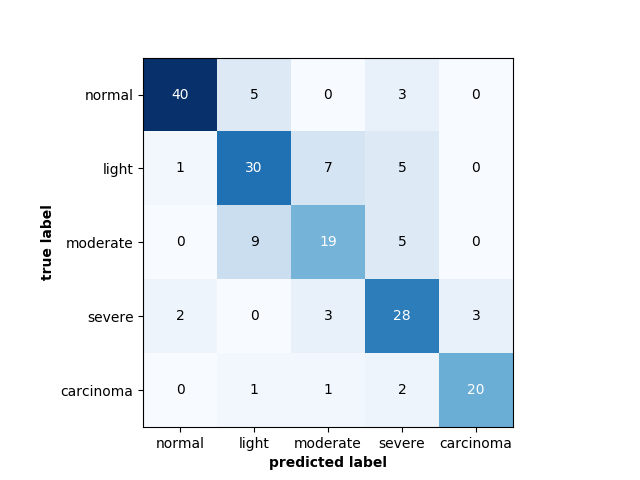}
\caption{Confusion Matrix given by Resnet-101 {Classifier + Regressor} on Herlev Severity Test Set}
\label{herlevclassregconfmat}
\end{figure}

On Figure \ref{herlevclassregdistrib} and Figure \ref{herlevclassregconfmat}, we can see that our Resnet-101, Classifier +
Regressor, makes less misclassifications than the classifier and lower MSE than
the regressor. Thus, we have an architecture performing on classification task
(mean AUC = 0.94) and on scoring severity task (average MSE = 0.51). What is
particularly appreciated here is that the 'extreme' classes ('normal' and
'carcinoma in situ') have the best AUC (respectively 0.98 and 0.97). The overall
accuracy of the order of 74.5\% and the binary accuracy was 94\% with an F1
score of 0.96.

\subsubsection{Pipeline Comparisons}

\begin{figure}[!t]
\centering
\includegraphics[width=\columnwidth]{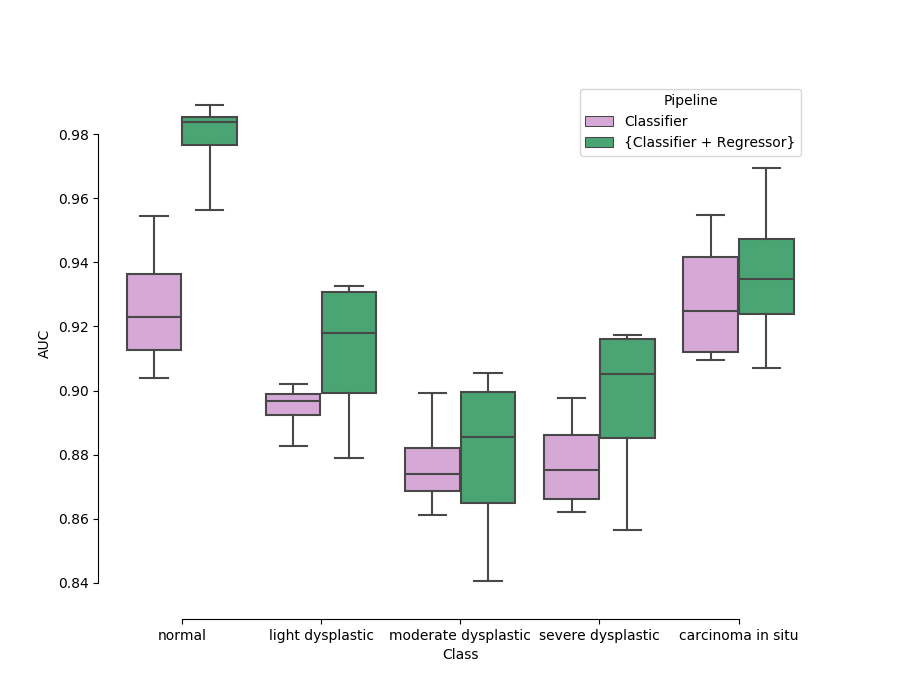}
\caption{AUC Distribution for Classifier \& {Classifier + Regressor} Pipeline Comparison using 4 random folds}
\label{aucboxplot}
\end{figure}

Figure \ref{aucboxplot} shows the AUC distribution per class obtained training
the classifier pipeline and the {classifier + regressor} pipeline on 4 random
folds. It brings to the fore how the regression loss does not change much
on the light dysplastic, moderate dysplastic and severe dysplastic classes
but improves 'extreme' cases especially 'normal' samples that were really
impacted by the ressemblance between 'normal columnar' and 'carcinoma in situ'
samples.

\section{Explainability / Interpretability}
\label{expsec}

Understanding how our trained model predicts the severity of new cells 
is an important step in validating its use. We need a method that provides meaningful
explanations, which ideally are related to the cytomorphological features and
used by cytotechnians and doctors during day-to-day routine. Gradient based
methods give the attribution to the classification associated with each input
feature given to the model, in the case of digital images of cytology slides,
the image pixels. This allows us to identify
and localized regions that contribute to the severity of the diagnosis. Integrated
Gradient \cite{Sundararajan17} is of particular interest due to its model
agnosticity and its baseline comparision.

In this section we are going to use an attribution method to understand what has
been learned by our models and on what cytomorphological features it relies to
assign a degree of malignancy. The Bethesda guidelines \cite{Bethesda01} states
that the main
cytomorphological features used to determine the severity are mostly based on
nucleus, we thus would expect the attribution to be in the nucleus region.

\paragraph{Integrated Gradient}

For the attribution we utilize a model agonistic methods, the integrated
gradient. As with most attribution methods it relies on the comparision between
the image and baseline (that is representative of the
absence of the class of the image) and computation of the gradient to the image.
The attribution map, $Am(x_i;F,x^{\prime})$ for an image $x$ gives the contribution
of
$i-\textrm{th}$ pixel given a model $F$ and baseline image $x^{\prime}$, $Am(x_i;F,x^{\prime}) =
 (x_i - x^{\prime}_i) . \sum_{k=0}^{m} \frac{\delta F(x^{\prime} + \frac{k}{m} . (x-x^{\prime}))}{\delta x_i} . \frac{1}{m}.$

\paragraph{Baseline Design}

What we are interested in here is how our model predicts the malignancy (i.e.
regression result), this is why we will try the Integrated Gradient method on
malignant samples i.e. dysplastic and carcinoma in situ samples. An
obvious abscence of object in Pap tests context is a white image (since
background of pap smears slides is white).

\paragraph{Qualitative Results}

\begin{figure}[!t]
\centering
\includegraphics[width=\columnwidth]{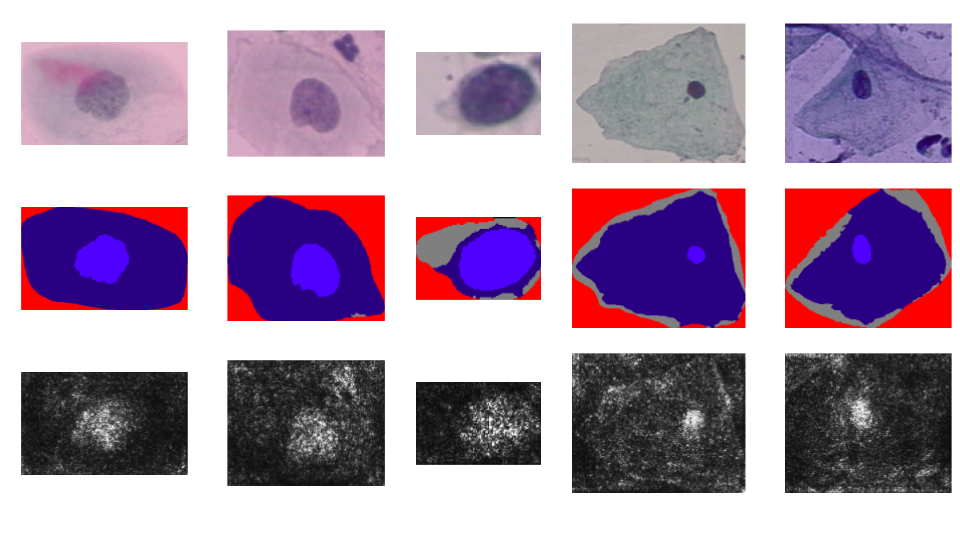}
\caption{Integrated Gradient Result on 5 images Herlev set (top) using a white baseline and cytological feature masks present in the dataset (middle). Attribution maps are shown showing "activated" pixels mostly in nucleus (bottom).}
\label{expqualitativeresults}
\end{figure}

Figure \ref{expqualitativeresults} shows examples of the attribution map
from integrated gradient method, along with the annotated cytology features of the
associated the images. This highlights that the malignancy scoring seem to be mainly
due to the nucleus.

\paragraph{Quantitative Results}

Here we make use of the annotation masks present in the Herlev set to create
specific attribution metrics. Given their role in the different consensus and
guidelines, we measure the amount of attribution within the nucleus and
cytoplasm compared to total attribution (respectively denoted as $At_{N}$ and
$At_{C}$), these contributions are given by,

\begin{eqnarray}
\label{attribution}
At_{N} &=& \frac{ \sum_{i \subset \mathcal{N}} Am(x_i)}{ \sum_{i} Am(x_i) }, \\
At_{C} &=& \frac{ \sum_{i \subset \mathcal{C}} Am(x_{i})}{ \sum_{i} Am(x_i) },
\end{eqnarray}
where $\mathcal{N}$ and $\mathcal{C}$, refer to the nucleus and cytoplasm pixels
 respectively and $Am$ is the attribution map defined before. In order to
 understand how much each region contributes to the model's prediction, we also
 compute the ratio of nucleus and cytoplasm attribution.

\begin{figure}[!t]
\centering
\includegraphics[width=\columnwidth]{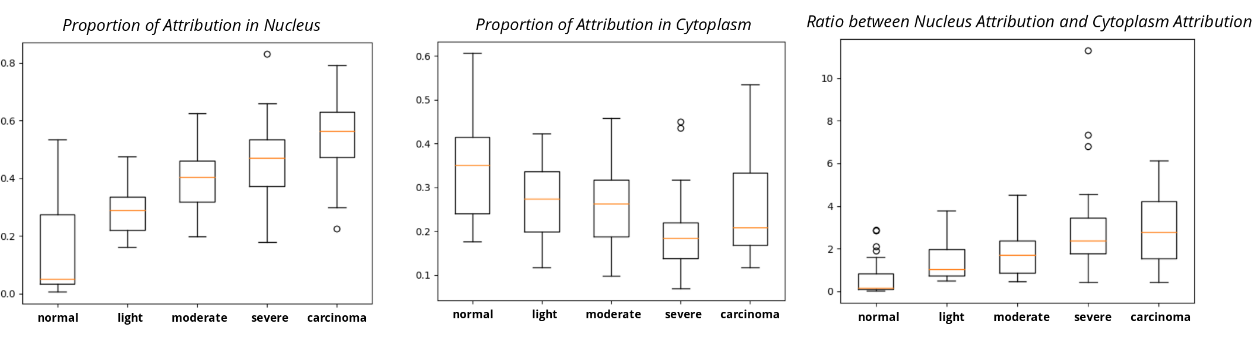}
\caption{Integrated Gradient Result on Herlev Test set using a white baseline and Associate Mask}
\label{expquantitativeresults}
\end{figure}

Figure \ref{expquantitativeresults} shows the distribution for $At_{N}$,
$At_{C}$, and their ratio for each severity class. It emphasises the
relevance of the nucleus over the cytoplasm for the model as the severity
increases. Particularly, in the case of carcinoma in situ, the nucleus
contributes $2$ times more than when classifying a normal case.

\section{Conclusion}

In this work, we have shown that a proper loss design, based on the
final goal of the medical exam under study, one can construct a model
that differentiates properly between normal and abnormal cells reaching a
severity accuracy of 74.5\%, a binary accuracy of 96.7\% was achieved along wiht a F1 score of 0.95.
Furthermore, we adapted an attribution method that can be used by doctors
to check the relevance of the network's decision. These two
contributions are essential in the construction of an automatic
diagnostic assistance method that can be trusted and accepted by
doctors.

\renewcommand{\abstractname}{Acknowledgements}
\begin{abstract}
We are grateful to our colleagues at Keen Eye and Telecom ParisTech (LTCI) for many valuable discussions, in particular Isabelle Bloch for advice and encouragment.
This work was supported by ANRT.
\end{abstract}


\begin{thebibliography}{99}

\bibitem{WHO06}
World Health Organization
\emph{Comprehensive cervical cancer control: a guide to essential practice}.
2006.

\bibitem{Bar15}
Yaniv ~Bar and Idit ~Diamant and Lior ~Wolf and Hayit ~Greenspan
\emph{Deep learning with non-medical training used for chest pathology identification}.
Medical Imaging 2015: Computer-Aided Diagnosis, 2015.

\bibitem{Ronneberger15}
Olaf ~Ronneberger and Philipp ~Fischer and Thomas ~Brox
\emph{U-Net: Convolutional Networks for Biomedical Image Segmentation}.
Medical Image Computing and Computer-Assisted Intervention – MICCAI 2015.

\bibitem{Krizhevsky12}
Alex ~Krizhevsky and Ilya ~Sutskever and Geoffrey E. ~Hinton
\emph{ImageNet Classification with Deep Convolutional Neural Networks}.
Advances in Neural Information Processing Systems 25, 2012.

\bibitem{He15}
Kaiming ~He and Xiangyu ~Zhang and Shaoqing ~Ren and Jian ~Sun
\emph{Deep Residual Learning for Image Recognition}.
2016 IEEE Conference on Computer Vision and Pattern Recognition (CVPR).

\bibitem{Lu17}
Le ~Lu and Ling ~Zhang and ~al
\emph{DeepPap: Deep Convolutional Networks for Cervical Cell Classification}.
IEEE Journal of Biomedical and Health Informatics, 2017.

\bibitem{Forslid17}
G. ~Forslid and H. ~Wieslander and ~al
\emph{Deep Convolutional Neural Networks for Detecting Cellular Changes Due to Malignancy}.
IEEE International Conference on Computer Vision Workshops, 2017

\bibitem{Sundararajan17}
Mukund ~Sundararajan and Ankur ~Taly and ~al
\emph{Axiomatic Attribution for Deep Networks}.
ICML 2017.

\bibitem{Bethesda01}
D. ~Solomon and D. ~Davey and ~al
\emph{The 2001 Bethesda System : Terminology for Reporting Results of Cervical Cytology}.
2001

\bibitem{Deng09}
J. ~Deng and W. ~Dong and ~al
\emph{A Large-Scale Hierarchical Image Database}.
CVPR 2009.

\bibitem{Hyeon17}
Jonghwan ~Hyeon and Ho-Jin ~Choi and ~al
\emph{Diagnosing cervical cell images using pre-trained convolutional neural network as feature extractor}.
IEEE International Conference on Big Data and Smart Computing, 2017

\bibitem{Bora16}
Kangkana ~Bora and Manish ~Chowdhury and ~al
\emph{Pap Smear Image Classification Using Convolutional Neural Network}.
Proceedings of the Tenth Indian Conference on Computer Vision, Graphics and Image Processing, 2016

\end{thebibliography}
\end{document}